\documentclass[aps,prl,reprint,superscriptaddress]{revtex4-1}

\usepackage{amsmath}
\usepackage{amsfonts}
\usepackage{amssymb}
\usepackage{enumerate}
\usepackage{amscd}
\usepackage{amsthm}
\usepackage[english]{babel}
\usepackage[pdftex]{graphicx}
\usepackage[pdftex]{thumbpdf}
\usepackage[bookmarks=true,pdfpagemode=UseOutlines]{hyperref}

\begin{document}
\title{Mott transition in the A15 phase of Cs$
_{3} $C$_{60}$: absence of pseudogap and charge order}
\author{H.~Alloul}
\affiliation{Laboratoire de Physique des Solides, CNRS, Univ. Paris-Sud, Universit\'{e} Paris-Saclay, 91405 Orsay, France}
\author{P.~Wzietek}
\affiliation{Laboratoire de Physique des Solides, CNRS, Univ. Paris-Sud, Universit\'{e} Paris-Saclay, 91405 Orsay, France}
\author{T. Mito}
\affiliation{Laboratoire de Physique des Solides, CNRS, Univ. Paris-Sud, Universit\'{e} Paris-Saclay, 91405 Orsay, France}
\author{D.~Pontiroli}
\affiliation{CNISM and Dipartimento di Fisica, Universit{\`{a}} di Parma - Via G.P.Usberti 7/a, 43100 Parma, Italy}
\author{M.~Aramini}
\affiliation{Department of Physics, University of Helsinki, Gustaf Hällströmin katu
2a, P.O. Box 64 00014 Helsinki, Finland}
\affiliation{CNISM and Dipartimento di Fisica, Universit{\`{a}} di Parma - Via G.P.Usberti 7/a, 43100 Parma, Italy}
\author{M.~Ricc{\`{o}}} 
\affiliation{CNISM and Dipartimento di Fisica, Universit{\`{a}} di Parma - Via G.P.Usberti 7/a, 43100 Parma, Italy}
\author{J.P. Itie}
\affiliation{Synchrotron SOLEIL, Orme des Merisiers, Saint-Aubin, BP 48, 91192 Gif-sur-Yvette Cedex, France}
\author{E. Elkaim}
\affiliation{Synchrotron SOLEIL, Orme des Merisiers, Saint-Aubin, BP 48, 91192 Gif-sur-Yvette Cedex, France}
\date{\today}

\begin{abstract}
 
We present a detailed NMR study of the insulator to metal transition induced by an
applied pressure $p$ in the A15 phase of Cs$_{3}$C$_{60}$. We evidence that the insulating
antiferromagnetic (AF) and superconducting (SC) phases only coexist in a narrow $p$ range. At fixed $p$, in the metallic state above the SC transition $T_c$, the $^{133}$Cs and $^{13}$C NMR spin lattice
relaxation data are seemingly governed by a pseudogap like feature. We prove that this feature, also seen in the $^{133}$Cs NMR shift data, is rather a signature of the Mott transition which broadens and smears out progressively for increasing $(p,T)$. The analysis of the variation of the quadrupole splitting $\nu _{Q}$ of the $^{133}$Cs NMR spectrum 
precludes any cell symmetry change at the Mott transition and only monitors a weak variation of lattice parameter. These results open an opportunity to consider theoretically the Mott transition in a multiorbital three dimensional system well beyond its critical point. 

\end{abstract}

\maketitle

In the cuprates a Mott AF insulating state is driven into a SC state by a chemically induced increase of carrier content. Another paradigm for correlated electron systems is to induce such a change of electronic state by an applied pressure $p$ \cite{Mott68,Mottbook}. An experimental realization requires a material with a large on-site Coulomb repulsion $U$ to stabilize an insulating state at ambient $p$. A significant compressibility is then required to permit an increase with $p$ of the bandwith $W$, allowing to span a large range of $U/W$ values. Due to these restrictions only few compounds exhibit such an ideal Mott transition.

 The most investigated compounds so far have been the vanadium oxides which  
exhibit a metal to insulator transition (MIT) above room $T$, though towards non SC ground states. As for many other cases studied a change in the atomic structure often occurs at the MIT, as for instance in doped VO$_{2}$
which displays a spin-Peierls transition concomitant to the MIT\cite{Pouget75}. On the contrary V$_{2}$O$_{3}$ undergoes a MIT \cite{McWhan69prl1,McWhan69prl2} which has been considered as a prototypical case since the 1960s.
It has been mostly studied in samples with slight Cr or Ti substitution on the V site that induces doping but also an uncontrolled
amount of disorder. This yields an inhomogeneous phase coexistence range
near the MIT \cite{Leonov2015}.

Recently original Mott transitions towards a SC ground state have been revealed in organic materials either of the BEDT-TTF family \cite{Lefebvre00prl,Kagawa05nmat} or in the
fulleride compound Cs$_{3}$C$_{60}$ \cite{Takabayashi-Science323,Ihara-PRL104}. Such compounds are highly compressible as the bindings between the organic or fullerene molecules occur through Van der Waals interactions. In the case of Cs$_{3}$C$_{60}$ detailed studies of the electronic
properties have been limited as actual samples are air
sensitive powders, often multiphased. However NMR experiments permitted one to establish that a MIT occurs in both isomeric cubic forms of this compound. The A15 phase exhibits a transition
from a N\'{e}el AF to a SC metal  with increasing $p$ while the fcc phase evolves from a frustrated magnetic state to a SC state\cite{Ihara-PRL104,Jeglic-PRB80}.
Recent low $T$ experiments on a well controlled nearly A15-pure sample permitted us \cite{Wzietek2014} a 
thorough study of the SC state properties on the high $p$ side of the Mott transition.

On this improved sample we evidence here  
that the AF-SC phase coexistence range at the Mott transition 
is rather small. We furthermore show that at fixed $p$ the thermal lattice expansion drives a transition between paramagnetic metal (PM) and insulating (PI) states, as reported for the organic 
conductor \cite{Kagawa05nmat} and as was suggested \cite{Brouet-PRB66bis} for the fulleride compound 
Na$_{2}$CsC$_{60}$. From NMR data taken on multiphased samples we did initially suggest that such a transition occurs as well in the fcc-Cs$_{3}$C$_{60}$ \cite{Ihara-EPL94}. Here the much higher experimental accuracy permits us to reveal the absence of pseudogap, which differentiates the fullerides from the cuprates near their respective Mott transitions.

We further demonstrate that, at variance with the organic compounds, the slope $dp/dT$ of the MIT remains positive over the whole phase diagram. 
Finally the $^{133}$Cs NMR quadrupolar splitting $\nu _{Q}$ 
only reveals a small step variation  of the lattice parameter at the Mott transition. The increase at high pressure of the apparent  width of the transition 
 suggests a possible location of the critical point $p_{c}$ in the phase diagram and allows us to compare the continuation of the Mott transition to a Widom line \cite{Brazhkin2011,Xu2005pnas,Xu2014prl}. These results should also allow one to consider theoretically the incidence of multiorbital \cite{Nomura2016}  and Jahn Teller \cite{Klupp2012nat,Zadike2015} effects detected in these compounds.

\paragraph{Mott transition and low $T$ AF-SC coexistence.}
We present here a detailed investigation of the $T$ variation of the $^{133}$Cs NMR spectra at low $p$ in the pure Mott-AF phase and for $p > p_{c0}\sim 5$ kbar for which the SC state is fully established \cite{Wzietek2014}.
The $^{133}$Cs NMR spectra exhibit a similar quadrupole splitting (for $I=7/2$) in both paramagnetic regimes but their shifts and width differ in the AF and SC states\cite[sec.I]{sup}. 
\nocite{Nelson,Piermarini75,Ganin-NMAT7,Fujiki-PRB02,Burkhart96}
 This permitted us to evaluate at a given $p$ the AF phase content from the signal loss occuring in the AF state in a specific frequency window of the NMR spectrum, as detailed in \cite[sec.I]{sup}.
As displayed in Fig.~1(a) the loss onsets below $T_{N}\sim 47 $ K 
for all pressures $p<5.4$ kbar while it only onsets at $T_{c}\sim 35 $ K for $p\geqslant 5.9$ kbar. Therefore the magnetic and metallic states only coexist on a narrow $p$ range with a quasi constant N\'eel temperature.
The comparison of the \textit{local} NMR data on the magnetic state with the SC fraction taken on the \textit{macroscopic} diamagnetic data is displayed in Fig.~1(b). It establishes that, up to $T=50K$, the transition occurs at the nearly $T$ independent pressure $p_{c0}= 5.1 \pm 0.3$kbar.  The sharpness of this transition with a small $p$ extension $\Delta p/p_{c}=\pm 6\%$  of the coexistence regime is a good evidence for the expected first order character of the Mott transition. 
\begin{figure}[tb]
\begin{center}
\includegraphics[width=0.99\hsize]{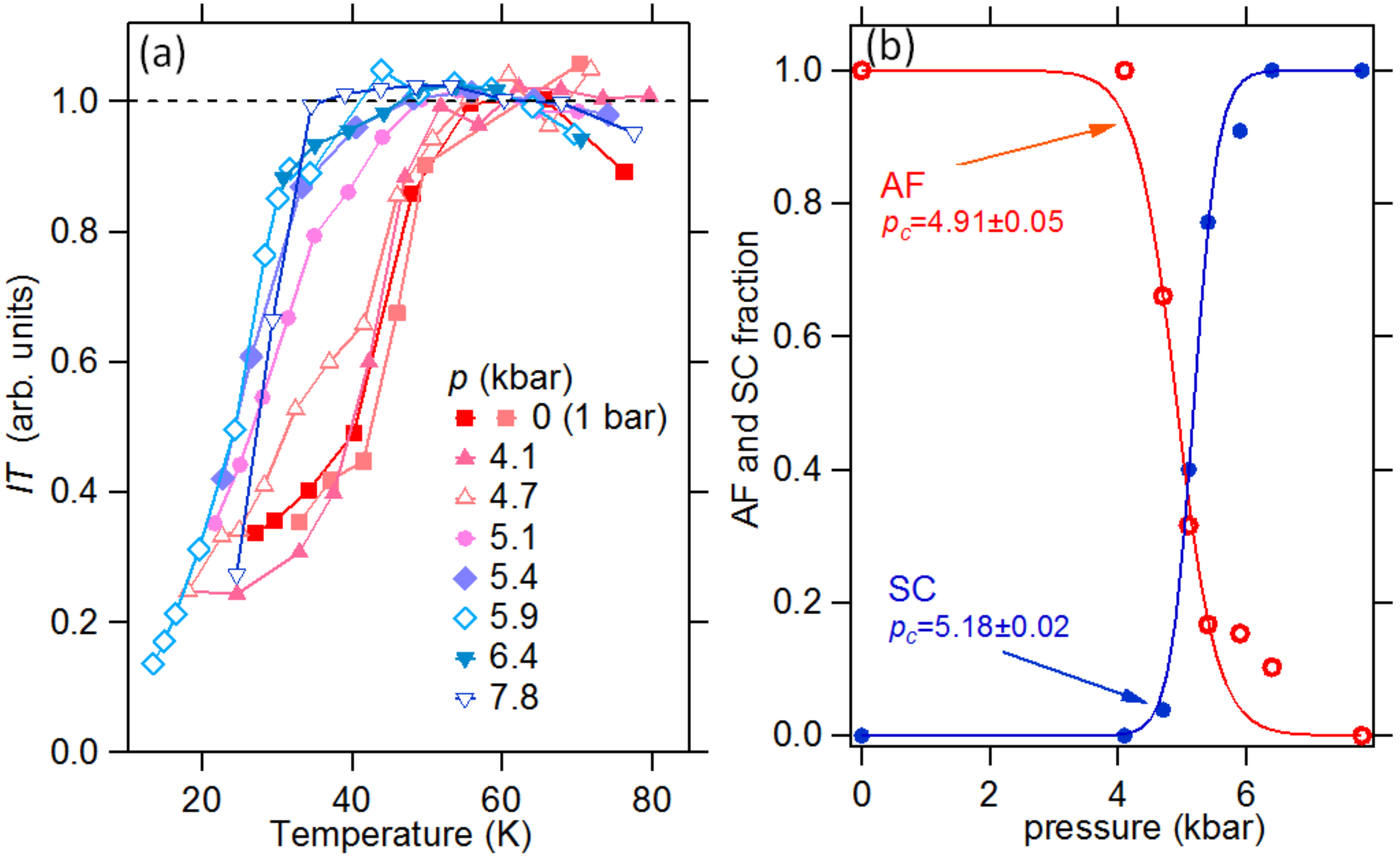}
\end{center}
\caption{\textbf{(a)} The  $^{133}$Cs NMR intensity corrected for its Curie $1/T$ dependence in the spectral range delineated in Fig.~S1(a) of \cite{sup}. For any $p$ the intensity loss due to the AF contribution is seen to onset below 50 K. \textbf{(b)} AF fraction obtained by integrating the intensity betwen 50 K and 35 K. Comparison with the SC fractions obtained in \cite{Wzietek2014} from diamagnetic data.}
\label{Fig1} 
\end{figure}
\paragraph{Pseudogap-like variation of the NMR spin-lattice relaxation $T_{1}$ in the metallic state.} To follow the evolution of the electronic properties toward high $T$ in the metallic state, 
we have first taken $T_{1}$ data on both $^{13}$C and $^{133}$Cs for a chosen pressure of 5.9 kbar for which the sample is fully in the metallic state at low $T$. In the paramagnetic state, the nuclear magnetization recovery was found to exhibit a stretched exponential behaviour $M(t)/M_{0} = exp[-(t/T_{1})^\beta]$, with an identical exponent value $\beta = 0.85(5)$  ($\beta$ did only slightly vary below $T_{N}$ and $T_{c}$ 
\cite{Wzietek2014}). The large decrease of $(T_{1}T)^{-1}$ due to the opening of the SC gap 
is clearly evidenced in Fig. 2(a). 
There, one can see as well a large step-like increase for $T>80$ K which is found to scale quite well for the two nuclear spin species. Being here very near to the Mott transition, one might consider that, in analogy to the case of underdoped cuprates, an incomplete pseudogap \cite{Alloul89prl} opens and is followed by a full SC gap opening at lower $T$. However, for a wave-vector-dependent pseudogap due to AF fluctuations, distinct $T$ dependences of $T_{1}$ should occur for the two nuclear spin species as $^{13}$C resides on the magnetic C$_{60}$ molecules (cf. $^{63}$Cu in cuprates \cite{Walsteadt,Takigawa91prb}) while $^{133}$Cs, being located in a symmetric position in the AF lattice, would hardly sense the AF fluctuations (cf. $^{89}$Y or $^{17}$O in cuprates \cite{Alloul89prl}). 
\paragraph{Recovery of the insulating state at high $T$.} Such a pseudogap being ruled out by these data, we shall see hereafter that $^{133}$Cs $T_{1}$ and NMR shift data for a series of pressures above and below $p_{c0}$ do allow us to establish that the high-$T$ transition corresponds to the restoration of an insulating paramagnetic (PI) state. 
\paragraph{(i) Paramagnetic state spin-lattice relaxation.}
In Fig. 2(b) we report first former 1 bar data \cite{Ihara-PRL104} taken on a large $T$ range on distinct samples sealed in 
glass capillaries. There we recognize the large increase of $(T_{1}T)^{-1 }$  with
decreasing $T$ \cite{Jeglic-PRB80, Ihara-PRL104}, followed by a sharp 
decrease due to the opening of the magnetic gap in the AF state below $T_{N}=47$ K.  In
the PI state this variation corresponds to a Curie-Weiss-like behaviour 
 $(T_{1}T)^{-1 }= B/(T+ \theta)$ with $\theta=60\pm5$ K, of the order of magnitude of $T_{N}$ as expected for a dense 
paramagnet \cite{Jeglic-PRB80, Ihara-PRL104}. 
 In Fig. 2(b) we report then a series of $(T_{1}T)^{-1}$ data for $p>2$ kbar taken on the sample housed in the pressure cell. Below 4.7 kbar they exhibit convex $T$ dependences similar  to that  seen  on the 1 bar samples. The small 20\% magnitude deviation could be mainly assigned to sample and NMR probe differences in the two experimental set-ups.   
For all pressures $p>5.1$ kbar, we see that the formerly studied \cite{Wzietek2014} SC state increase of $(T_{1}T)^{-1}$ up to $T_{c}$ is followed by a second step-like increase as that seen in  Fig.~2(a). This is evidenced by the inflexion point and the concave variation displayed by all the data curves. 
One might notice that at higher $T$ values, $(T_{1}T)^{-1}$ recovers a decreasing behaviour strikingly 
identical to that seen for the PI state at 4.1 kbar.
This ascertains that the step increase of $(T_{1}T)^{-1}$ indeed locates the MIT. 
For $p\geqslant 7.8$ kbar an intermediate plateau of $(T_{1}T)^{-1 }$ suggestive of a paramagnetic metal (PM) Korringa behaviour \cite{Abragam} can be noticed between the two regimes. 
\begin{figure}[tb]
\begin{center}
\includegraphics[width=0.99\hsize]{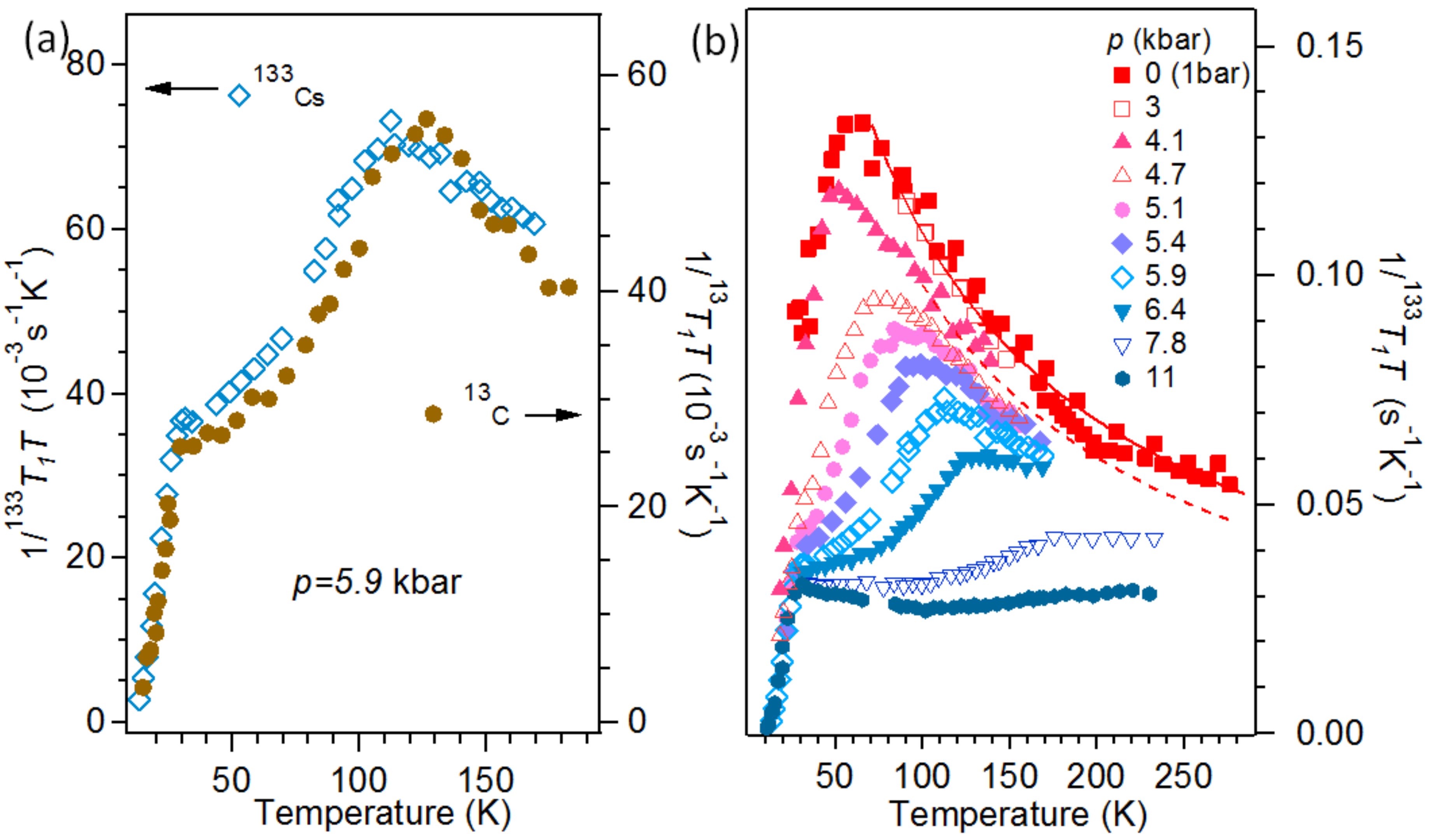}
\end{center}
\caption{\textbf{(a)} $(T_{1}T)^{-1}$ for $^{13}$C and $^{133}$Cs taken at
5.9 kbar with the vertical axes scaled. 
\textbf{(b)} The $^{133}(T_{1}T)^{-1}$ data are plotted versus $T$ for a 
series of pressures $p$. The Curie-Weiss fits for 1bar and 4.1 kbar (see text) are plotted as full and dotted lines respectively. The step associated with the MIT shifts towards higher $T$ with increasing $p$.} 
\label{Fig2}
\end{figure}

\paragraph{(ii) NMR shifts from the $^{133}$Cs NMR spectra.}
The transition from a PM to a PI can be detected as well on the electronic spin susceptibility which can be monitored from the $^{133}$Cs NMR shift.
With a single $I=7/2$ Cs nuclear spin site, the spectra could be fitted as exemplified in \cite[sec.II]{sup} to deduce the NMR shift tensor, the quadrupole frequency $\nu _{Q}$ and its asymmetry parameter $\eta$. For all pressures below 11 kbar the isotropic contribution $K_{iso}$ to the NMR shift increases with $T$ as shown in Fig. 3(a). In the PI range, this increase above $T_{N}$ is due to the progressive decorrelation of the local moments. For $p\eqslantgtr 5.9$~kbar
the recovery of the  NMR shift at $T_{c}\gtrsim30$ K is followed by a step-like increase of $K_{iso}$ quite identical to that seen for the $(T_{1}T)^{-1}$ data. One can even see in \cite{sup} that $(T_{1}T)^{-1}$ linearly scales with $K_{iso}$ in the intermediate $T$ range where the two quantities increase sharply. These results allow us to establish unambiguously that the transition $T_{MIT}(p)$ from the PM to the PI state shifts towards higher $T$ for increasing $p$.  
\begin{figure}[tb]
\begin{center}
\includegraphics[width=0.99\hsize]{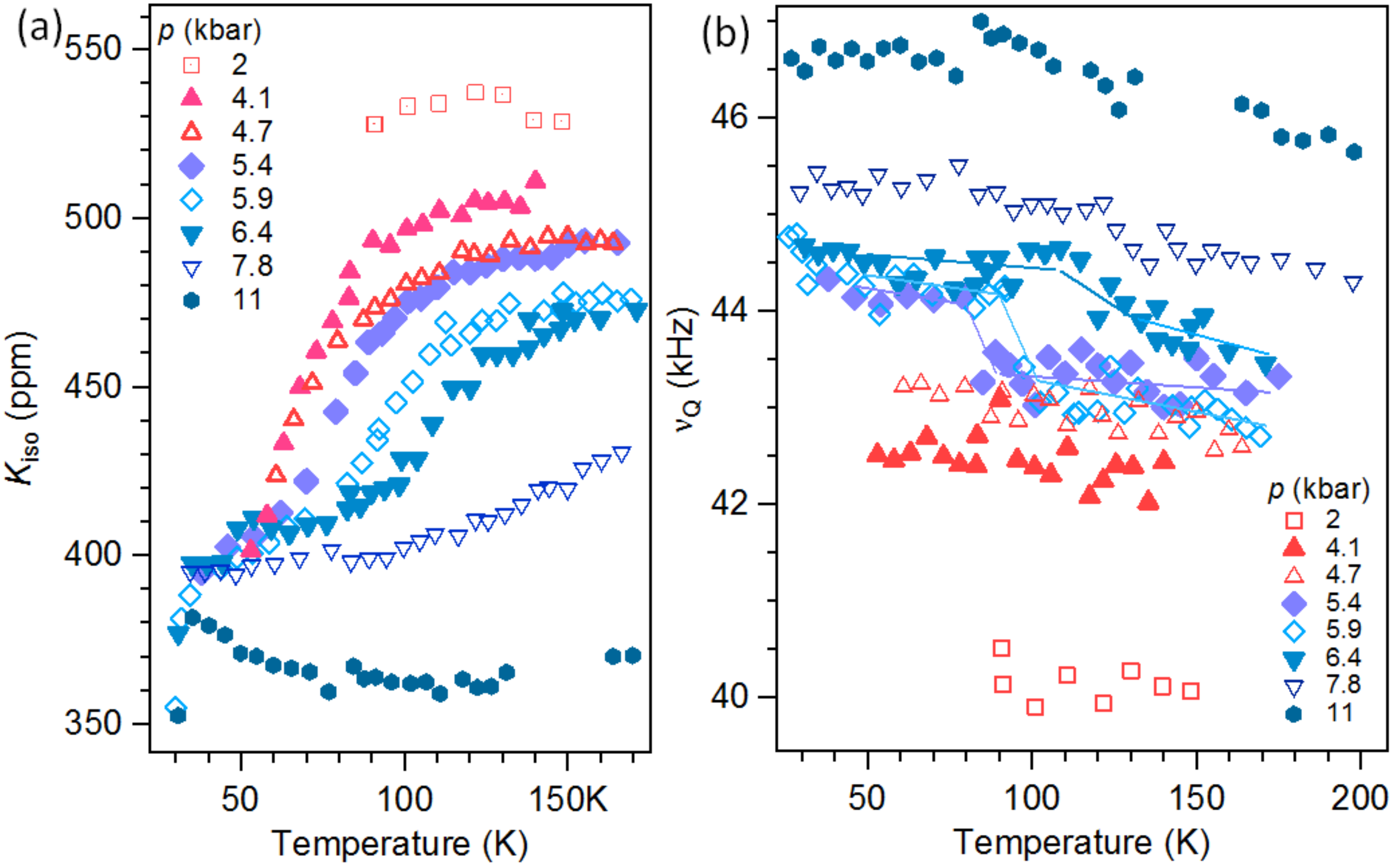}
\end{center}
\caption{The $^{133}$Cs NMR spectral analyses done in \cite[sec.II]{sup} yield: \textbf{(a)} $K_{iso}$ on a large ($T,p$) range which displays step variations for increasing $T$ similar to those found for $(T_{1}T)^{-1}$ in Fig. 2(b); \textbf{(b)} $\nu _{Q} $ data do reveal small jumps of $\sim 1$ kHz for $5.4<p<6.4$ kbar which disappear at higher $p$. Lines are guides to the eye.}
\label{Fig3}
\end{figure} 
\paragraph{Structural changes at the transition.}
Former synchrotron radiation x-ray data taken on a very large $p$ range only permitted to detect an overall variation of the lattice parameter \cite{Takabayashi-Science323}. We also took x-ray data, described in  \cite[sec.III]{sup}, which allows us to confirm that the A15 lattice cell symmetry is not subsequently modified from the Mott state up to $p=11$ kbar. 
The nuclear quadrupole frequency $\nu _{Q}$ deduced from the $^{133}$Cs NMR spectra is also directly related to the distribution of charges and then to the atomic structure and the ordering of the C$_{60}$ molecular orbitals. The extensive spectral fits described in \cite{sup} permitted us to determine the $T$ variations of $\nu _{Q} $ shown in Fig.3(b). Those are naturally found quite smooth in the absence of any crossing of the MIT for $p<4.7$ kbar. However for $5.1<p<7$ kbar minute step decreases of $\nu _{Q}$ with increasing $T$ are evidenced near the transition temperatures detected from the $T_{1}$ and $K_{iso}$ data of Fig.2(b) and 3(a). No such signature of the transition could be detected at higher $p$.
 We recall that $\nu _{Q}$ gives a measure of the electric field gradient (EFG) on the Cs atomic site with $\nu _{Q}\propto  d^{2}V/dr^{2}$ where $V(r)$ is the potential around the Cs site located at $r=0$. As $V(r)$ has dimension $L^{-1}$, $\nu _{Q}$ has dimension $L^{-3}$. Assuming a uniform lattice contraction with increasing $p$ this dimensional analysis demonstrates that one roughly expects $\nu _{Q}\propto a^{-3}=1/v$, where $v$ is the lattice unit cell volume. We shall consider first the qualitative information gained on the phase diagram from the $\nu _{Q}$ data while a quantitative analysis of the variation of lattice parameter will be attempted later.
\paragraph{Phase Diagram.}
The data presented here-above permitted us to evidence on $(T_{1}T)^{-1}$, $K_{iso}$ and $\nu _{Q}$ that the metallic state crosses over toward the insulating state with increasing $T$ at a given $p$. We are now able to discuss the main characteristics of the transition line $T_{MIT}(p)$ and its apparent width.
\paragraph{(i) Transition line.} We already know from the discussion done on the AF-SC fractions that the transition line is nearly vertical at $p=5.1$ kbar, from $T=0$ to 50 K. For higher $p$, the temperature $T_{Q}$ at which a small step decrease of $\nu _{Q}$ is detected permits to locate the evolution of the transition up to $p\sim 6.4$ kbar. The magnetic responses $(T_{1}T)^{-1}$ and $K_{iso}$, which scale with eachother in the metallic regime (see \cite[sec.II]{sup}), give a determination of the transition which can be followed at larger $p$. The maximum of $(T_{1}T)^{-1}$ marks the full restoration of the insulating state that is the  upper limit $T_{max}$ for the MIT. One might as well assign a width to the transition and consider that the MIT sits at the inflexion point $T_{mid(T_1)}$ of $(T_{1}T)^{-1}$ versus $T$ (or equivalently $T_{mid(K)}$ for $K_{iso}$ $vs$ $T$) which can be obtained by fitting the fixed $p$ data with a step-like function with a half intensity width $\Delta_{T}$, as done in \cite[sec.II]{sup}.  The data for $T_{max}$, $T_{Q}$, $T_{mid(T_1)}$ and $T_{mid(K)}$  plotted in Fig.4 give therefore a good representation  of
the variation of the MIT, which bends over for increasing $p$ in the actual phase diagram.
In the fcc-Cs$_{3}$C$_{60}$  phase a similar MIT transition line has been suggested solely from NMR $T_1$ data taken  with  increasing pressure (Fig. 5 and 6 of \cite{Ihara-EPL94}) or for Rb-substituted samples \cite{Zadike2015}.   
 
\paragraph{(ii)  Transition width and critical point.}
We may anticipate that in the absence of sample defects and of $p$ distribution in the pressure cell, the transition should be much narrower 
and display the hysteresis expected
for a first order transition, as found in organics \cite{Kagawa05nmat}. If due to a purely inhomogeneous broadening, the 0.3 kbar half width of the transition determined for $p\simeq5$ kbar in Fig. 1 should increase linearly up to 0.5 kbar for $p=7.8$ kbar. From the 30 K/kbar slope of $T_{MIT}(p)$ of Fig. 4, one would expect at most $\Delta_{T} =15$ K obviously much smaller than $\Delta_{T}=50$ K obtained experimentally at 7.8 kbar \cite[sec.II]{sup}. Therefore, the large increase of $\Delta_{T}$ at high $p$ is apparently governed by a fundamental intrinsic effect. This behaviour is indicative of a suppression of the first order transition above a critical pressure $p_{c}$. The latter can be tentatively located at $\simeq7$ kbar as the transition cannot be detected on $\nu _{Q}$ data between 6.4 and 7.8 kbar. Therefore beyond $p_{c}$ the transition line is rather the loci of a continuation of the first order transition usually considered as a Widom line \cite{Brazhkin2011,Xu2005pnas,Xu2014prl}. Its location is well known to depend somewhat on the thermodynamic quantity employed, even for classical systems such as the liquid-gas transition. 
\begin{figure}[tb]
\begin{center}
\includegraphics[width=0.99\hsize]{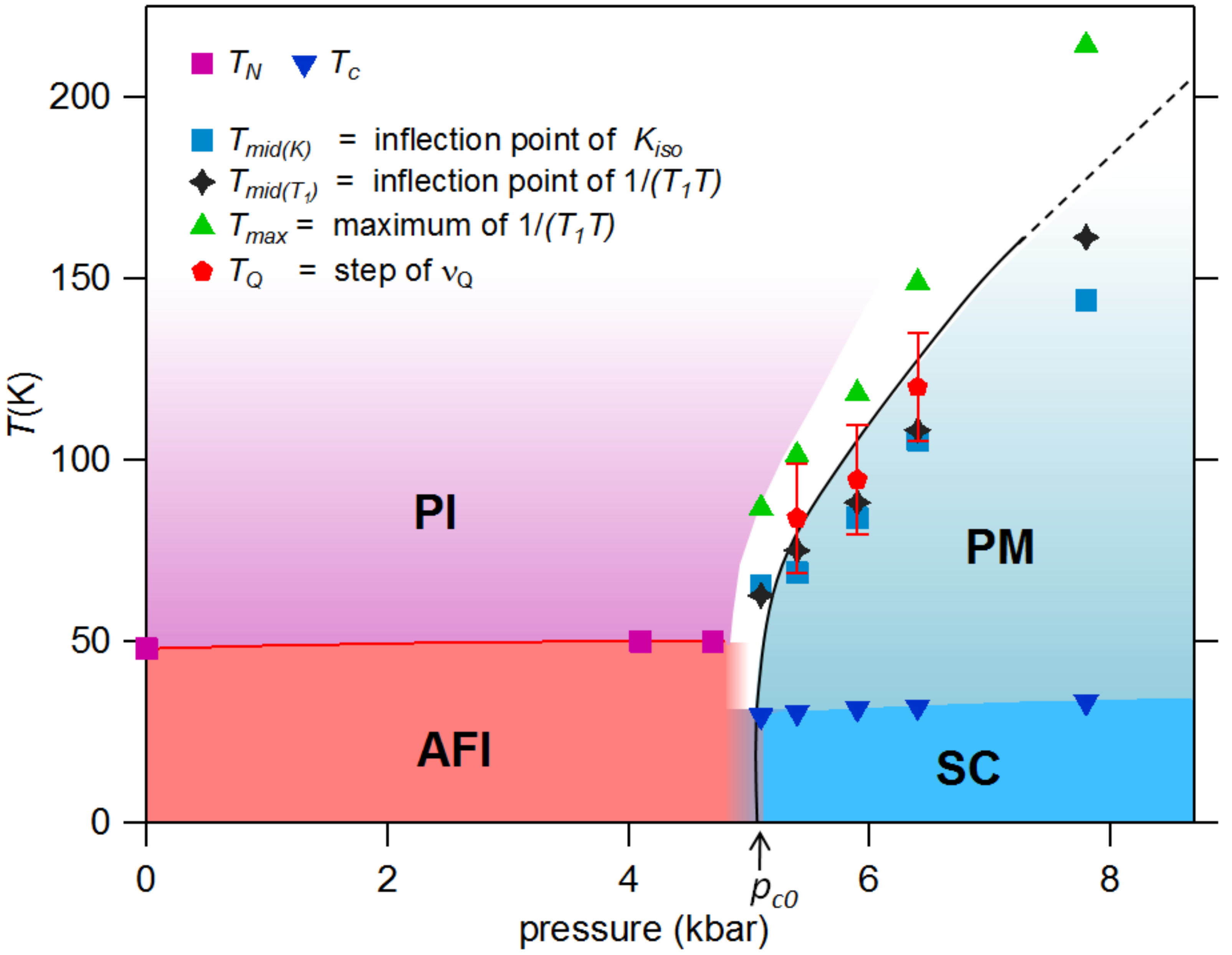}
\end{center}
\caption{ MIT transition temperatures $T_{Q}$, $T_{mid(T_1)}$, $T_{mid(K)}$  and $T_{max}$ deduced from  $^{133}$Cs NMR data for $p>5.4$ kbar. The $T_{Q}$ data mark the weak first-order regime (full line) while $T_{max}$ delineates the apparent upper half width of the transition given as the blank zone. This suggests that the transition becomes a crossover (dotted line) at high $(T,p)$ beyond a critical point $p_{c}\simeq7$ kbar (see text). The AF and SC fractions measured in Fig.1(b) yielded $p_{c0}=5.1\pm 0.3$ kbar below 35 K. The phase diagram is completed by the horizontal $T_{N}$ line below 5.1 kbar, and the $T_{c}$ dome above 5.1 kbar.}
\label{Fig4}
\end{figure}
\paragraph{Discussion of $\nu _{Q}$.}
Let us consider the physical significance of the $p$ and $T$ variations of $\nu _{Q}$ on the Cs site. As discussed hereabove, for a uniform contraction of the unit cell in a simple point charge model one expects $d\nu _{Q}/\nu _{Q}=-dv/v$. The x-ray data \cite[sec.III]{sup} taken at $T$ =14.5 K correspond to an initial linear $p$ dependence of the lattice cell volume $dv/(vdp) = -6.2\times10^{-3}/$~kbar. From Fig. 3(b) we obtain $d\nu _{Q}/\nu_{Q}dp=0.02 /$ kbar for $p<5$ kbar at a fixed  $T$ of 50 (or 80) K for which the sample remains in the PI state. Therefore $\nu _{Q}$ exhibits a three times faster increase with unit cell contraction than expected. While this limits the significance of the model, 
we may consider that this experimental scaling can be applied as well for the $\simeq 1$ kHz decrease of $\nu _{Q}$ at the MIT. That would correspond to about $\delta v/v\simeq 0.008$, that is an increase of lattice parameter $ da/a\simeq 2\times10^{-3}$, which is within the error bar of the synchrotron x-ray data we have collected so far in diamond anvil cells \cite{sup}.

Let us recall here that $\nu _{Q}$ involves two contributions $\nu _{Q}= \nu _{Q,i}+ \nu _{Q,b}$. The first one, associated with the EFG created by large distance ions is correctly estimated within the above model. The second contribution is due to the ionic charges and electrons distributed in the Cs first-nearest neighbours molecular C$_{60}$ orbitals. In the  absence of modification of the C$_{60}$ ball diameter with $p$ and $T$ (confirmed in \cite[sec.III]{sup}) the unit cell does not contract uniformly. This might essentially control the local contribution $\nu _{Q,b}$ associated with the C$_{60}$ molecular orbitals, which could as well be sensitive to the appearance of extended states at the Mott transition. 
\paragraph{Conclusion.}
We have shown here that A15-Cs$_{3}$C$_{60}$ presents a quasi ideal Mott transition and performed 
an NMR study which allowed us to reveal the absence of pseudogap and charge order near the Mott transition. This lack of competing order might explain that $T_c$ remains quite high up to the MIT, conversely to the case of hole doped cuprates.
We could establish altogether the evolution of the MIT first order line in the $(p,T)$ phase diagram towards a continuous transition. Quite remarkably, we also find that the insulating state restored from the metallic state by thermal expansion of the lattice has magnetic properties quite analogous to those found at high $T$ in the pure insulating Mott state. This might indicate that quantum fluctuations do govern the continuous transition at high $(p,T)$, which therefore differs from the Widom line occuring in classical systems e.g. the liquid-gas transition. Let us recall as well that the maximum of $(T_{1}T)^{-1}$ versus $T$ which has been revealed for long \cite{Holczer93epl} in the dense A$_{3}$C$_{60}$ light-alkali compounds is quite similar to that detected here above $p=7.8$ kbar. This gives a strong hint that in all these compounds the maximum of $(T_{1}T)^{-1}$ represents  the crossover towards the local moment behaviour of the PI state.    
Finally, contrary to the case of the organic triangular lattice crystals, the critical line has a positive slope with respect to $p$. 
This entropy related slope, which suggests higher entropy in the insulating state linked with orbital degeneracy, 
qualitatively agrees with the results of DMFT calculations for fcc A$_{3}$C$_{60}$ (A = K, Rb, Cs) \cite{Nomura2016}. 
More quantitative simulations taking into account the  A15 A$_{3}$C$_{60}$  lattice geometry could permit to investigate the interplay between orbital and spin degrees of freedom in this genuine Mott transition.

\begin{acknowledgments}
We acknowledge SOLEIL for provision of synchrotron radiation facilities and would like to thank S. Ravy  for assistance in taking X-ray data on the CRISTAL beamline (projects 2010286 and 20110678).
 We also have had stimulating discusions with  S. Biermann, Y. Nomura, V. Dobrosavlevic and M. Rozenberg. 
\end{acknowledgments}

%\bibliographystyle{apsrev4-1}
%\bibliography{a15mott}

%merlin.mbs apsrev4-1.bst 2010-07-25 4.21a (PWD, AO, DPC) hacked
%Control: key (0)
%Control: author (72) initials jnrlst
%Control: editor formatted (1) identically to author
%Control: production of article title (-1) disabled
%Control: page (0) single
%Control: year (1) truncated
%Control: production of eprint (0) enabled
%

\end{document}